\documentclass[10pt,a4paper]{article}
\usepackage{f1000_styles}
\usepackage{float}
\usepackage{cite}

\newcommand{\fref}[1] {Figure~\ref{#1}}

\begin{document}
\pagestyle{fancy}

\title{Threshold studies for a hot beam superradiant laser including an atomic guiding potential}
\author{Martin Fasser$^{1}$}
\author{Christoph Hotter$^{1}$}
\author{David Plankensteiner$^{1}$}
\author{Helmut Ritsch$^{1}$}
\affil{
$^1$Institut f\"ur Theoretische Physik, Universit\"at Innsbruck, Technikerstr. 21a, A-6020 Innsbruck, Austria}

\maketitle
\thispagestyle{fancy}

Corresponding author: Martin Fasser (\href{mailto:martin.fasser@uibk.ac.at}{martin.fasser@uibk.ac.at})\\
\\
\\
\begin{abstract}
\\
\textbf{Background:} \\
Motivated by the outstanding short time stability and reliable continuous operation properties of microwave clock masers, intense worldwide efforts target the first implementation of their optical analogues based on narrow optical clock transitions and using laser cooled dilute atomic gases. While as a central line of research large efforts are devoted to create a suitably dense continuous ultracold and optically inverted atom beam source, recent theoretical predictions hint at an alternative implementation using a filtered thermal beam at much higher density. Corresponding numerical studies give encouraging results but the required very high densities are sensitive to beam collimation errors and inhomogeneous shifts. Here we present extensive numerical studies of threshold conditions and the predicted output power of such a superradiant laser involving realistic particle numbers and velocities along the cavity axis. Detailed studies target the threshold scaling as a function of temperature as well as the influence of eliminating the hottest part of the atomic distribution via velocity filtering and the benefits of additional atomic beam guiding. Using a cumulant expansion approach allows us to quantify the significance of atom-atom and atom-field correlations in such configurations. \\    
\textbf{Methods:}  \\
To enable studies with realistic high particle number we implement a simulation framework based on a first order as well as a second-order cumulant expansion of the coupled atom-field dynamics with the atomic center of mass motion represented by classical trajectories. We assume a velocity filtered initial thermal distribution and guiding forces from prescribed optical potentials along the cavity axis, while the transverse atomic motion is not explicitly included as dynamical variable. Effectively, atomic motion along the cavity axis induces a time varying atom-cavity coupling determined by the cavity mode structure and the optical guiding potentials. Our model includes a continuous effective incoherent pump mechanism. Using such simulations we study how the intra-cavity photon number depends on the atom number, initial velocity distribution and atomic guiding. 
Comparing the results in different expansion orders including customized mixed order models allows to extract the relative importance of atom-atom and atom-field correlations in different operating regimes.\\
\textbf{Results:} \\
We predict necessary conditions to achieve a certain threshold photon number depending on the atomic temperature and density. In particular, we show that the temperature threshold can be significantly increased by using more atoms. Interestingly, a velocity filter removing very fast atoms has only almost negligible influence despite their phase perturbing properties. On the positive side an additional conservative optical guiding towards cavity mode antinodes leads to significantly lower threshold and higher average photon number. Interestingly we see that higher order atom-field and direct atom-atom quantum correlations play only a minor role in the laser dynamics, which is a bit surprising in the superradiant regime.\\
\textbf{Conclusions:} \\
A hot atomic beam laser operated in the superradiant regime can achieve useful power levels, if the inhomogeneous atomic broadening from the velocity distribution along the cavity axis can be compensated by using sufficiently more atoms. Velocity filtering to remove very hot atoms negatively influences the average photon number but can potentially reduce power fluctuations as well as a reduced linewidth and cavity pulling effects. Adding a confining optical lattice potential, however, creates significant modifications of the dynamics, especially for higher temperatures where our model simulations predict significantly higher laser output power.
\end{abstract}

\clearpage
\pagestyle{fancy}

\section*{Introduction}
Ever since the demonstration of the superior operation of an optical atomic clock~\cite{takamoto2005optical, bloom2014optical, ludlow2015optical} with respect to microwave implementations people started wondering about a possible active clock implementation in form of a superradiant laser~\cite{Haake1993superradiant, meiser2009prospects, chen2009active, meiser2010steady, meiser2010intensity, bohnet2012steady, vuletic2012almostlightlesslaser, maier2014superradiant, Zhang2018montecarlo, debnath2018lasing, hotter2019superradiant, gogyan2020characterisation, zhang2021ultranarrow, Bychek2021superradiant, kazakov2022ultimate, kristensen2023subnatural, zhang2023development}. In analogy to the very successful hydrogen micro-maser technology~\cite{goldenberg1960atomic} such an active device in principle promises at least similar stability and accuracy at a reduced technical cost and fragility. Here an implementation based on a dilute gas of cold clock atoms within a high-Q optical cavity was predicted to be a very promising path to go~\cite{meiser2009prospects}. As a particular example, model setups based on a beam of inverted ultracold atoms traversing an optical resonator~\cite{kazakov2013active, kazakov2014active, tang2022prospects} were predicted to yield unprecedented stability and accuracy. However, as the operating optical wavelength is several orders of magnitude shorter than the typical maser lines, the resulting technological challenges are similarly larger. 

Already the construction of a suitable beam source is virtually equivalent to operating a continuous atom laser~\cite{salzburger2007atom-photon, holland1996theory} and thus poses a way more challenging task than its microwave analog. As the cavity mode volume is also way smaller the atoms have to be tightly confined before sending them through the resonator. Yet another problem is creating atomic inversion with minimal perturbation~\cite{hotter2022continuous}. Despite intense efforts which recently lead to amazing technical
advancements on guided atom beams~\cite{bennets2017steads, chen2019continuous, Escudero2021steady, chen2022continuous} and pulsed superradiant lasing~\cite{norcia2016superradiance, norcia2016cold, norcia2018frequency, laske2019pulse, schaffer2020lasing, tang2021cavity, hotter2023cavity, bohr2023collectively}, a fully continuous operating device~\cite{bohnet2012steady, kristensen2023subnatural} has not been realized. 

In recent calculations it was pointed out that a high precision could already be achieved using a much hotter atomic ensemble with a sufficiently high intra-cavity density~\cite{liu2020rugged, jaeger2021regular, jaeger2021superradiant, tang2022prospects}. Here the sheer number of contributing atoms allows to cross the laser threshold inducing an efficient collective synchronization process towards a very narrow effective linewidth. This approach at least in principle requires significantly less components and would allow for a rather compact setup involving only a beam oven, an optical resonator and a pump laser to create inversion. The precise conditions for a concrete setup, however, cannot be reliably analytically predicted and thus require large scale numerical simulations~\cite{tang2022prospects}. Here we set up a framework for such simulations which can deal with realistically large atom numbers and velocities. 
This allows to identify the minimal operating conditions and study the influence of extra elements as velocity filters or guiding potentials. 


\section*{Model}
\label{sec:model}
\begin{figure}[ht]
\includegraphics[width=0.9\textwidth]{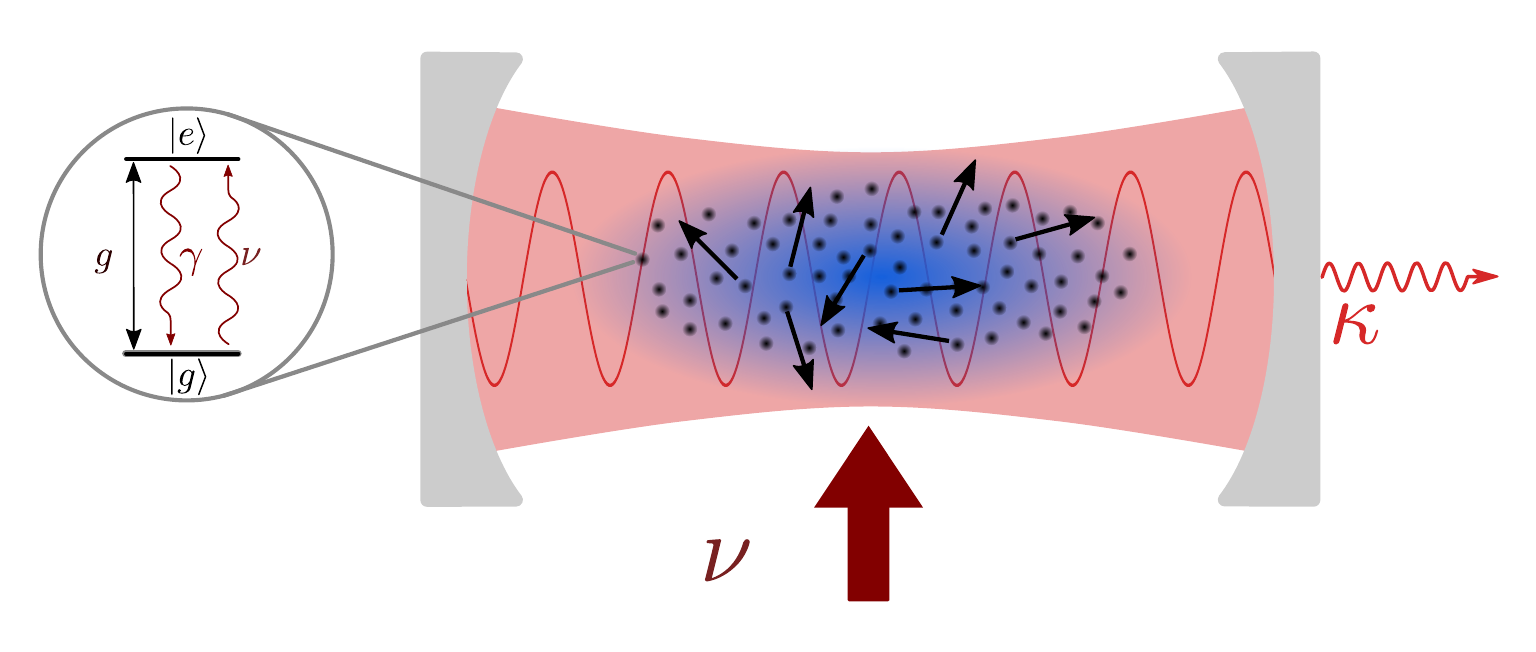}
\caption{\emph{Schematic of the system.} Atoms are incoherently pumped with rate $\nu$ and couple to an optical cavity. The clusters move according to a Gaussian velocity distribution with standard deviation $\sigma_\mathrm{v}$, which relates to a certain temperature $T$. This motion is included along the cavity axis with a position dependent coupling strength $g$. The atoms decay with the rate $\gamma$ and the photons leak out of the cavity with a rate $\kappa$.}
\label{fig:model}
\end{figure}

We consider an ensemble of $N$ two-level atoms with a transition frequency of $\omega_\text{a}$ coupled to a single mode cavity with resonance frequency $\omega_\text{c}$, see~\fref{fig:model}. The coherent dynamics of this system is governed by the Tavis Cummings Hamiltonian, which in the rotating frame of the atomic transition frequency
$\omega_\text{a}$ reads
\begin{equation}
H = \Delta a^\dagger a + \sum^N_{j=1} g(x_j) (a^\dagger \sigma^\mathrm{ge}_j + \sigma^\mathrm{eg}_j a),
\label{eq:H_TC}
\end{equation}
with the atomic transition operator of the $j$-th atom $\sigma^\mathrm{ge}_j = | g \rangle_j \langle e |$, the cavity photon creation (annihilation) operator $a^\dagger$ ($a$), and the cavity-atom detuning $\Delta = \omega_\text{c} - \omega_\text{a}$. Atoms at a position $x$ along the cavity axis couple to the cavity according to the mode function $g \cdot f(x) = g \cdot \cos(2 \pi x / \lambda)$, where $g$ is the maximal cavity coupling strength and $\lambda$ the wavelength of the cavity photons. The motion of the atoms is treated classically by an initial atomic velocity drawn from a Gaussian distribution with width $\sigma_v$ corresponding to a temperature $T$. In the beginning we treat freely moving atoms, later on we investigate the influence of an external optical potential. In both cases the atomic motion is fully determined by the initial velocity and position. 

Additional to the coherent dynamics we need to account for the dissipative processes, which are described by Liouvillian term $\mathcal{L}[\rho]$ in the master equation for the system density matrix $\rho$ 
\begin{equation}
\dot{\rho} = i [\rho, H] + \mathcal{L}[\rho].
\label{eq:master_eq}
\end{equation}
In the Born-Markov approximation the Liouvillian super-operator for a jump operator $J_i$ with a corresponding rate $\Gamma_i$ can be written in Lindblad form as
\begin{equation}
\mathcal{L}[\rho] = \sum_i \frac{\Gamma_i}{2} \left( 2 J_i \rho J_i^\dagger - J_i^\dagger J_i \rho - \rho J_i^\dagger J_i \right).
\label{eq:lindblad}
\end{equation}
The considered dissipative processes are listed in table~\ref{tab:dissipative}, including cavity photon losses, individual spontaneous decay as well as incoherent pumping of the two-level atoms.
\begin{table}[b]
\caption{\emph{Dissipative Processes.} The system features a damped cavity mode, atomic decay and an incoherent pump.}
\begin{center}
\begin{tabular}{ lccl }
 \hline
 $i$ & $c_i$ & $\Gamma_i$ & Description \\
 \hline
 1 & $a$ & $\kappa$ & cavity photon losses \\
 2 & $\sigma_j^\mathrm{ge}$ & $\gamma$ & decay from $|2\rangle_j$ to $|1\rangle_j$ \\
 3 & $\sigma_j^\mathrm{eg}$ & $\nu$ & excitation of the $j$-th atom \\
\hline
\end{tabular}
\end{center}
\label{tab:dissipative}
\end{table}

As we want to simulate systems with up to one million atoms, we cannot solve the full master equation due to exponential scaling of the size of the Hilbert space with the particle number. To this end we employ a mean field approach for operator averages. For a not explicitly time-dependent operator $\mathcal{O}$ we can calculate the time-evolution of its expectation value as
\begin{equation}
\frac{\mathrm{d}}{\mathrm{d}t}\langle \mathcal{O} \rangle = \mathrm{Tr} \{ \mathcal{O} \dot{\rho}\} = i \langle [H, \mathcal{O}] \rangle + \langle \mathcal{\bar{L}} [\mathcal{O}] \rangle
\label{eq:expect_o}
\end{equation}
with 
\begin{equation}
\mathcal{\bar{L}} [\mathcal{O}] = \sum_i \frac{\Gamma_i}{2} \left( 2 J_i^\dagger \mathcal{O} J_i - J_i^\dagger J_i \mathcal{O} - \mathcal{O} J_i^\dagger J_i \right),
\end{equation}
where we inserted the master equation \eqref{eq:master_eq} into Eq. \eqref{eq:expect_o} and used the cyclic property of the trace. To truncate the set of equations we approximate higher order quantum correlations with the cumulant expansion method~\cite{Kubo62, plankensteiner2021quantumcumulantsjl}. 

The first order approximation (mean field) is sufficient to accurately describe our system (see Appendix), the corresponding set of equations is
\begin{subequations} 
\label{eq.meanfield}
\begin{align} 
\dot{\langle a \rangle }&= -\Bigl( \frac{\kappa}{2}+i \Delta \Bigl) \langle a \rangle -  i g \sum_{{m}} f(x_{{m}})\langle \sigma^\mathrm{ge}_{{m}} \rangle \\
\dot{\langle \sigma^\mathrm{ge}_{{m}} \rangle}&=-\frac{\gamma+\nu}{2}\langle \sigma^ \mathrm{ge}_{{m}} \rangle +2ig f(x_{{m}})(\langle \sigma^\mathrm{ee}_{{m}} \rangle- \frac{1}{2})\langle a \rangle \\
\dot{\langle \sigma^\mathrm{ee}_{{m}} \rangle}&=-(\gamma	+ \nu)\langle \sigma^\mathrm{ee}_{{m}} \rangle+ig f(x_{{m}})(\langle a \rangle^{*}\langle \sigma^ \mathrm{ge}_{{m}} \rangle-\langle \sigma^ \mathrm{eg}_{{m}} \rangle^*\langle a \rangle)+ \nu,
\end{align}
\end{subequations}
where the last two equations exist for each atom from $m=1$ to $N$. The number of equations in this approximation scales linearly with the atom number $N$. Still, to simulate a system with several hundred thousand atoms we need further approximations. To this end, we assume that groups of $K$ atoms behave in a sufficiently identical way, such that we can describe them with the same equations. This means, we group atoms into clusters. For the set of equations in~\ref{eq.meanfield}, this means that the index $m$ now runs only from $m=1$ to the number of clusters $N_\mathrm{Cl}$, but the source term $-i g \sum_{{m}} f(x_{{m}})\langle \sigma^\mathrm{ge}_{{m}} \rangle$ in Eq. (\ref{eq.meanfield}a) gets enhanced by the clustersize $K$. In the following, we use a clusternumber of $N_\mathrm{Cl}=400$, which is a good compromise between simulating the real setup and still having a feasible computation time (see Appendix). 

\section*{Photon number dependence on temperature}
\begin{figure}[ht]
\includegraphics[width=1.0\columnwidth]{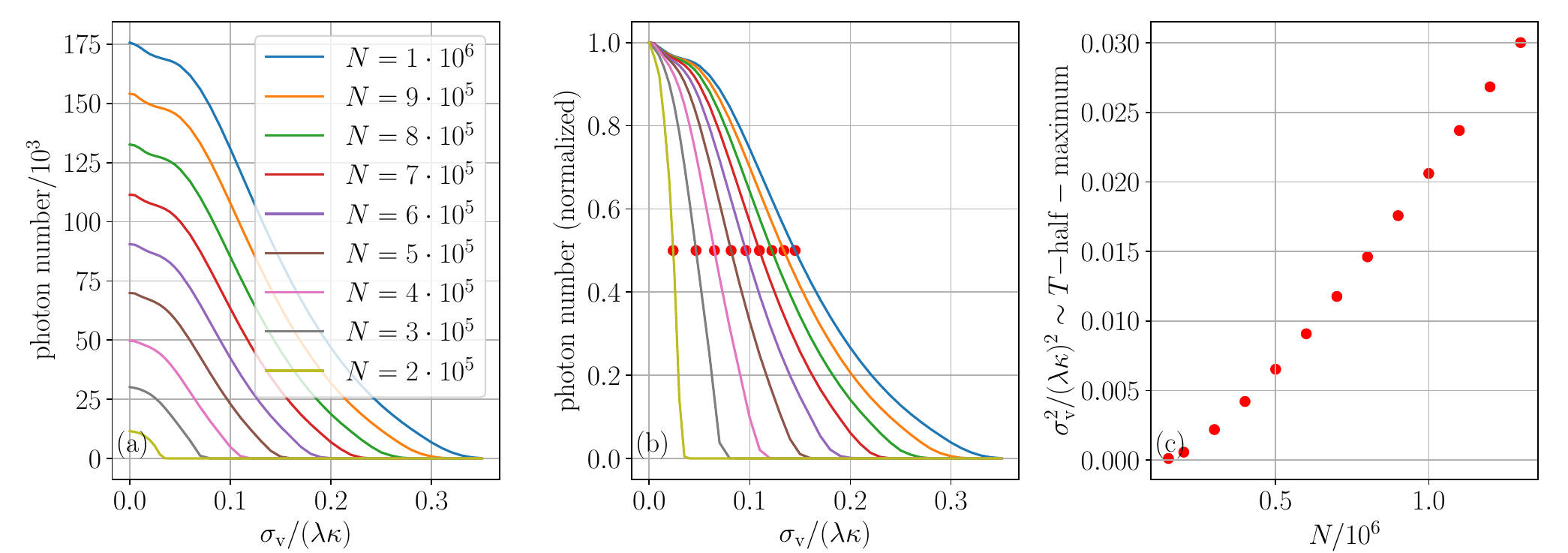}
\caption{\emph{Photon number dependence on temperature and atom number.} (a) steady state photon number (mean field, see Appendix for comparison to higher orders) as a function of the standard deviation of the velocity distribution $\sigma_\mathrm{v}$ for different atom numbers. (b) Photon number normalized to $\sigma_\mathrm{v}=0$ (maximum). The red dots indicate the value at which the curves have dropped to half of their initial value. After normalizing, the curves do not longer align, hinting to a nonlinear relation between atom and photon number. (c) Dependence of the threshold value of the velocity spread $\sigma_\mathrm{v}^2 \sim T$ on the number of atoms. Larger atom numbers allow for larger possible values of $\sigma_\mathrm{v}$, while maintaining a lot of photons inside the cavity. The parameters are $\nu=0.5$, $\Delta=0$, $g=0.00136$ and $\kappa=1$. We neglect $\gamma$ since it is much smaller than $\nu$, see Eq. \eqref{eq.meanfield}.}
\label{fig.fixg}
\end{figure}
First, we investigate the influence of the atom number $N$ and the temperature $T$ on the steady state photon number. The results in mean field (for a comparison to higher orders see Appendix) are depicted in~\fref{fig.fixg}. Unsurprisingly, we always end up with the highest number of photons for the lowest temperature. 
The drop-off of the photon number for higher temperature can be explained by the Doppler effect, as higher velocities of the individual atoms lead to them being shifted out of resonance. Comparing each curve to the others, we can investigate the significant influence of the atom number on the photon number: Higher atom numbers yield higher photon numbers. While this is not surprising, the crucial aspect of this result becomes apparent, when we normalize the curves to their photon number value for zero temperature (\fref{fig.fixg}(b)): The lacking coincidence of the curves hints to a nonlinear relation between atoms and photons. However, for low atom numbers we see the curves dropping to virtually 0 photons as early as for $\sigma_\mathrm{v}/(\lambda \kappa) \approx 0.05$, whereas for higher atom numbers such a temperature has hardly any effect on the photon number. For example, for $10^6$ atoms there are still approximately $95 \%$ of the photons at zero temperature in the cavity. The red dots indicate the points, at which the photon number has dropped to half their maximal value at zero temperature for each individual curve. These values are depicted in~\fref{fig.fixg}(c), they initially follow a quadratic relation, while for higher atom numbers the relation seems linear. From that graph we can clearly see, that using higher atom numbers allows a higher temperature while still having a decent amount of photons inside the cavity. This could be beneficial to experimental setups, where it is generally difficult to cool the atoms down to low temperature. However, it remains to be seen what this means to the linewidth of the outcoming photons, as higher temperature usually leads to Doppler broadening of the linewidth.
\section*{Adding a velocity filter}
\begin{figure}[ht]
\includegraphics[width=0.95\textwidth]{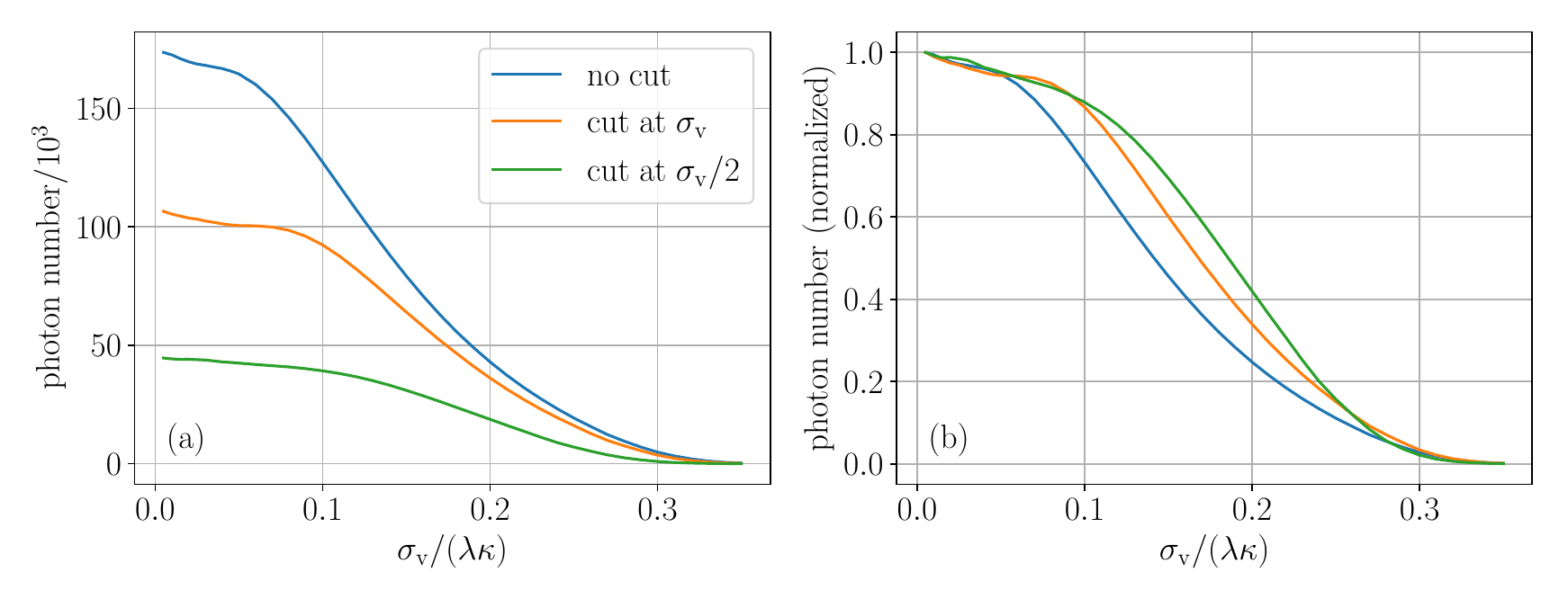}
\caption{\emph{Photon number for different velocity cutoffs.} (a) Mean steady state photon number. (b) Normalized to maximum photon number. Cutting off the the high velocity part of the distribution always yields a smaller steady state photon number, but it remains at a high level for increasing $\sigma_\mathrm{v}$. We only depict the results in mean field, as the results in mixed order almost perfectly align. The parameters are the same as in~\fref{fig.fixg}.} \label{fig.veldistcut}
\end{figure}
An interesting idea is to consider a cutoff in the velocity distribution for the atoms. Experimentally, this may be achieved by using velocity filter for the atomic beam to get rid of the atoms with very high velocity, creating a smaller but colder ensemble. This could be more advantageous than trying to narrow the velocity distribution by cooling, as cooling the atoms proposes an experimental challenge. To simulate such a filter, we start with the same number of atoms as before, but exclude all atoms from the dynamics with velocities higher than the cutoff. We depict the results in~\fref{fig.veldistcut} for a cutoff at $\sigma_\text{v}$ ($\approx 68 \%$ slowest atoms) and at  $\sigma_\text{v}/2$ ($\approx 38 \%$ slowest atoms). 
To compare, we also depict the previous results without cutoff. In~\fref{fig.veldistcut}(a) the steady state photon number is shown for the different cutoffs. By cutting off the high velocity part of the atomic distribution, we lose only fast atoms, but it always  leads to a decrease in photon number. A velocity filter therefore only reduces the photon number. It might still be useful with respect to the linewidth, which is not calculated at this point. 

However, the photon number seems to be more stable with increasing temperature, which can be seen in~\fref{fig.veldistcut}(b), where we normalize the graphs to their highest photon number (at zero temperature). While the photon number significantly drops already for low temperature (e.g. $\sigma_\mathrm{v} \approx 0.1$) for the uncut case, it stays relatively stable for the other cases, where we exclude the fast atoms.

In summary, including a velocity filter will negatively affect the photon number, but allow it to stay at a relatively stable value for low temperatures. However, it remains to be seen what effect such a velocity filter has on the linewidth of the atoms. In general, one would expect a velocity filter to be beneficial for the linewidth reduction, as the excluded, fast atoms increase the linewidth by means of the Doppler effect.
\section*{Dynamics including an optical lattice potential} 
\begin{figure}[ht]
\includegraphics[width=0.99\textwidth]{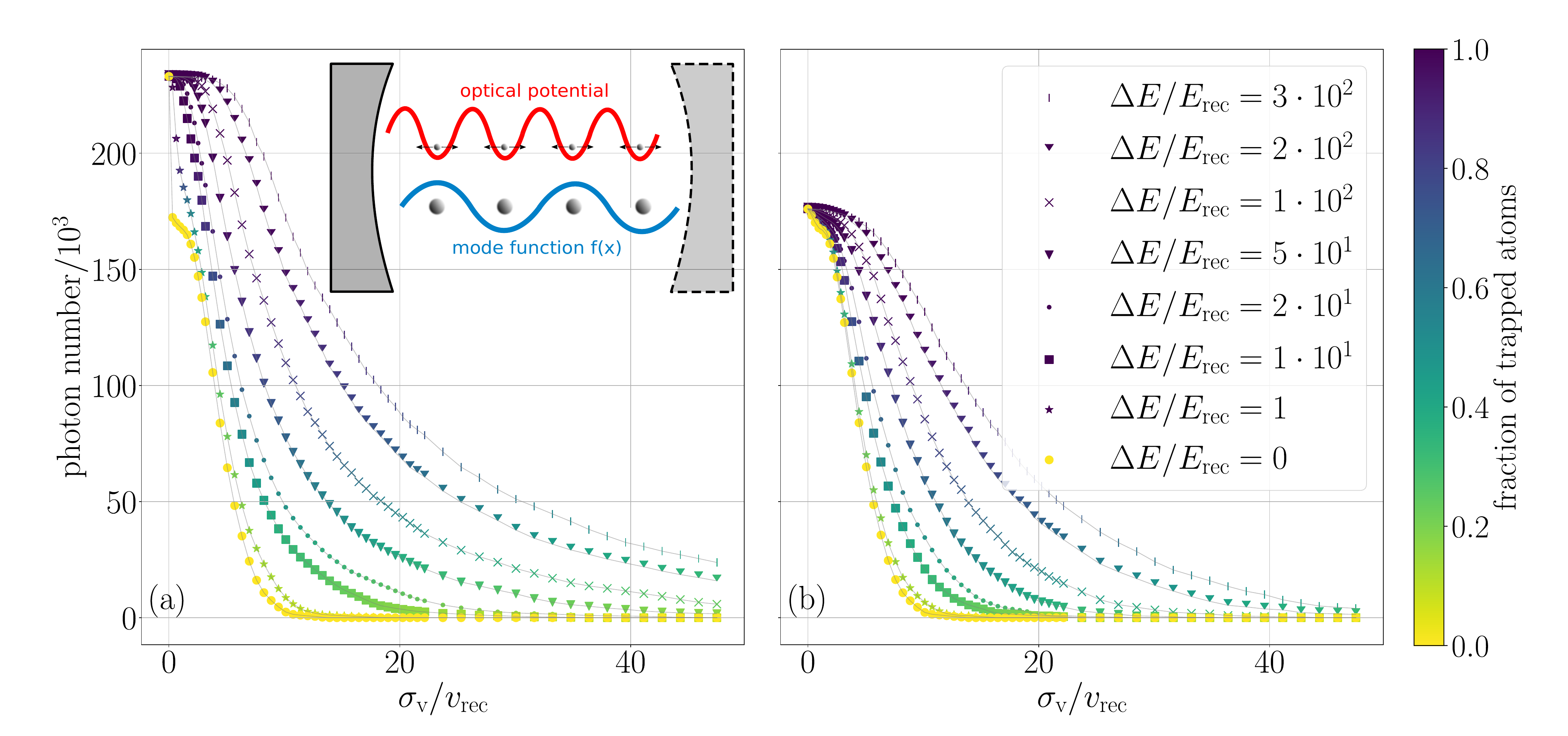}
\caption{\emph{Quasi-steady state photon number including a optical lattice potential with wavelength $\lambda_\mathrm{latt}=0.5\,\lambda$ (a) and $\lambda_\mathrm{latt}=0.99\,\lambda$ (b) with varying energy barriers.} The color of the data points indicates the fraction of trapped atoms, where at blue most of the atoms are trapped, while at yellow most of the atoms can escape their trap. (a) $\lambda_\mathrm{latt}=0.5\,\lambda$: Due to the choice of $\lambda_\mathrm{latt}$, for low temperatures most of the atoms are trapped at the antinode of the mode-function, leading to a higher atom-cavity coupling and higher photon number. For higher temperatures, most of the atoms escape their trap, but the photon number stays significantly above the barrierless case. This is due to the energy barriers, which slow the atoms down, yielding lower Doppler effect and therefore higher photon number. (b) $\lambda_\mathrm{latt}=0.99\,\lambda$: For low temperatures the results coincide, regardless of the chosen energy barrier. This means, that the atom-cavity coupling is the same as in the barrierless case, where we distributed the atoms equidistantly over a mode function. For higher temperatures, most of the atoms escape their trap, but the photon number behaves similarly to the case with $\lambda_\mathrm{latt}=0.5\,\lambda$. Once again, the atoms get slowed by the barrier, eventually leading to significantly higher photon number compared to the barrierless case. The parameters are the same as in~\fref{fig.fixg}.}
\label{fig.lattresult05}
\end{figure}

So far, we assumed that the atoms move freely without any potential. After some time at non-zero temperature, the atoms will be randomly distributed over the mode function. To compare that situation to the zero-temperature case with non-moving atoms, we always place the atoms equidistantly distributed over the mode function. Therefore, the differences in photon number stem from the movement of the atoms itself, not from the somewhat artificial location of atoms on the mode function.
The coupling between atoms and cavity field can be controlled (atleast for low temperature) by including an optical lattice potential, as the atoms are trapped at the potential minimums. Aligning these potential minimums with the antinodes of the mode function leads to enhanced coupling compared to the random sampling of the mode function in the case where atoms move freely. The potential $V(x)$ has the form $V(x)=\Delta E \cdot \cos^2(\frac{4 \pi}{\lambda_\mathrm{latt}} x)$ with spatial periodicity $\lambda_\mathrm{latt}$ and energy barrier $\Delta E$. By using the Hamiltonian formalism one ends up with two linear differential equations of first order for each cluster. These additional equations can simply be added to the equations in mean field (\ref{eq.meanfield}) for simulation purposes. To keep the computation time low, we now simulated only $N_\mathrm{Cl}=100$ clusters instead of 400 and only averaged over 10 trajectories instead of 50.\\
For the optimal case, we assume an optical potential with exactly half the wavelength of the mode function, that way the minima of the optical potential are situated at the maximum values of $|f(x_m)|$. The results of such a simulation are depicted in~\fref{fig.lattresult05}(a). To compare to the previous results, we include the graph that we already calculated, which is the result for an energy barrier of $\Delta	E=0$. The energy is expressed in terms of the recoil energy $E_\mathrm{rec}=\frac{\hbar^2 k^2}{2m}$, where $k=2 \pi / \lambda$, $v_\mathrm{rec}$ is the associated recoil velocity and $m$ is the mass of the atoms. To get exemplary values for these energy scales we use the $^{40}\mathrm{Ca}$ intercombination line between $^1\mathrm{S}_0$ and $^3\mathrm{P}_1$   with $\lambda=657\,\mathrm{nm}$ and $\Gamma = 2 \pi \cdot 375 \, \mathrm{Hz}$, the cavity linewidth is $\kappa=2 \pi \cdot 120 \, \mathrm{kHz}$.  To compare the two energy scales of the energy barrier and the velocity of the atoms, we color the data points according to the fraction of trapped atoms at each particular energy barrier and velocity. For high energy barriers and low velocities we have a lot of trapped atoms, these data points are colored blue, whereas for low energy barriers and high velocities almost no atoms are trapped, these points are colored yellow. As we can see from~\fref{fig.lattresult05}, the results change significantly by including a barrier. 
Let us first look at the case of low velocities/low temperatures: When the barrier is sufficiently high, then almost all the atoms are trapped at the antinodes of the mode function, leading to high coupling between atoms and cavities and eventually to higher photon number. For higher velocities/temperature, the number of photons stays significantly above the comparison graph for no energy barrier. On the one hand, this is because there are still atoms, that are trapped near the antinode of the mode function. On the other hand, there is a different effect at play, of which we can get a better insight when choosing the wavelength of the lattice slightly different from the wavelength of the mode function. Such a system may be of interest, as there usually is a mismatch between the effective optical lattice periodicity and the optical wavelength.\\
The results can be seen in~\fref{fig.lattresult05}(b), where we chose $\lambda_\mathrm{latt}=0.99\,\lambda$. Note, how the results converge for very low temperature, regardless of the value of the energy barrier $\Delta E$. This means, that the coupling is the same as in the barrierless case, where we distributed the atoms equidistantly over the mode function. So by choosing the wavelength of the optical lattice only slightly different to the wavelength of the mode function, we once again sample an average over the mode function, as long as the atoms sit on sufficiently many sites. Therefore the whole advantage of keeping atoms near the antinode of the mode function disappears and the coupling between the atoms and the cavity is the same compared to the coupling in the barrierless case. However, increasing the temperature, we still see a significant difference: Even though the average coupling is the same, the photon number is much higher. The atoms are decelerated, whenever they go over a potential barrier, yielding a reduced average velocity, which in turn decreases the Doppler effect and increases the photon number. 

In summary, for low velocities/temperatures, the change in photon number depends mostly on the change of coupling of the atoms, which again depends on the location of the minima of the optical potential compared to the antinodes of the mode function. For higher velocities/temperatures, the other effect of having a lower average velocity due to the barriers comes also into play. The lower average velocity leads to a decreased Doppler effect and therefore a significant increase of photon number.

\section*{Conclusions}
We studied incoherently pumped two-level-atoms inside a single mode cavity. In particular, we investigated the influence of temperature on the photon number. For high temperatures, where the atoms are fast, the photon number is suppressed because of the Doppler effect. However, we saw that increasing the atom number, to get sufficient collective atom-cavity-coupling, can compensate this suppression. 
Including a velocity filter leads to a lower amount of atoms participating in the dynamics and therefore, as the number of atoms is a significant parameter, to a lower photon number. We plan to calculate the spectrum to see if excluding the fast atoms has a decreasing effect on the linewidth.


With an optical lattice potential the coupling between the atoms and the cavity field can be controlled. For low temperatures (compared to the energy barrier), most atoms are trapped and the atom-cavity coupling (and in consequence the photon number) can be engineered. 
However, for high temperatures, even though most of the atoms are not trapped, the photon number significantly increases compared to the situation without a potential. In some cases, where virtually no photons are generated anymore for a certain $\sigma_\mathrm{v}$ without the potential, we still have a macroscopic photon occupation number, even if $\sigma_\mathrm{v}$ is 4 times as high, by including the optical potential. 
Therefore including an optical potential becomes very beneficial for higher temperatures.

\section*{Appendix. Equations in higher orders}
As mentioned in the main text, for our parameter regime the physics is accurately described by mean field. To verify this, we derive the equations for higher orders in the cumulant expansion and compare the numerical results with the mean field. For second order, one derives equations for two-operator averages. These will depend on other averages of two-operator and three-operator products. In order to close the set of differential equations one needs to derive equations for the two-operator averages and approximate the three-operator averages according to~\cite{Kubo62}
\begin{equation}
\label{eq:3opexpansion}
\langle abc \rangle \approx \langle ab \rangle \langle c \rangle + \langle ac \rangle \langle b \rangle +\langle cb \rangle \langle a \rangle - 2 \langle a \rangle \langle b \rangle \langle c \rangle.
\end{equation}
Using the phase invariance of the system~\cite{meiser2009prospects, plankensteiner2021quantumcumulantsjl} one ends up with the set of equations in second order of the cumulant expansion:
\begin{subequations} \label{eq.2ndorder}
\begin{align}
\dot{\langle a^{\dagger} a \rangle}=&- \Bigl( \kappa-2i \Delta \Bigl) \langle a^{\dagger} a \rangle +i g K \sum_{{m}} f(x_{{m}})(\langle a \sigma^\mathrm{eg}_{{m}}\rangle - \langle a^{\dagger}\sigma^\mathrm{ge}_{{m}}) \\
\dot{\langle a^\dagger \sigma^\mathrm{ge}_{{m}} \rangle} =&-\Bigl( \frac{\gamma+\kappa + \nu}{2} -i\Delta \Bigl) \langle a \sigma^\mathrm{eg}_{{m}} \rangle +ig f(x_{{m}}) (\langle \sigma^\mathrm{ee}_\mathrm{m} \rangle-\langle a^{\dagger}a \rangle) +ig \sum_{{j=1}}^{N} f(x_{{j}}) \langle \sigma^\mathrm{eg}_{{j}} \sigma^\mathrm{ge}_{{m}} \rangle \\[-10 pt]
&+2igf(x_{{m}}) \langle a^{\dagger}a \rangle \langle \sigma^{\mathrm{ee}}_{{m}}\rangle \nonumber \\[10 pt]
\dot{\langle \sigma^{\mathrm{ee}}_{{m}} \rangle} =&igf(x_{{m}})(\langle a^{\dagger}\sigma^\mathrm{ge}_{{m}} \rangle-\langle a \sigma^\mathrm{eg}_{{m}}\rangle) -(\gamma +\nu)\langle \sigma^{\mathrm{ee}}_{{m}} \rangle + \nu  \\[13 pt]
\dot{\langle \sigma^\mathrm{ge}_k \sigma^\mathrm{eg}_l \rangle}= &-( \nu + \gamma ) \langle \sigma^\mathrm{ge}_k \sigma^\mathrm{eg}_l \rangle - ig f(x_{{k}}) \langle a \sigma^\mathrm{eg}_l \rangle + ig f(x_{{l}}) \langle a \sigma^\mathrm{eg}_k \rangle - \nu \langle \sigma^\mathrm{ge}_k \sigma^\mathrm{eg}_l \rangle \\[6 pt]
&+2ig f(x_{{k}}) \langle \sigma^\mathrm{ee}_k \rangle \langle a \sigma^\mathrm{eg}_j \rangle -2ig f(x_{{l}}) \langle \sigma^\mathrm{ee}_j \rangle \langle a \sigma^\mathrm{eg}_k \rangle.
 \nonumber
\end{align}
\end{subequations}
The advantage of this second order is, that we can include atom-field and atom-atom correlations, however, we pay the price in form of a higher computation time, as the number of equations now scales quadratically with the cluster number instead of linear as above. The terms involving two atom operators $\langle \sigma_i \sigma_j \rangle$ are responsible for the quadratic scaling, as both $i$ and $j$ are between 1 and $N$. In an attempt to take the advantage of the upsides of both first order and second order, we approximate two-operator products if they both are atom operators by splitting them, but keep the correlations between atom and field, we call this particular truncation "mixed order":
\begin{subequations} \label{eq.mixedorder}
\begin{align}
\dot{\langle a \rangle}= &-\Bigl( \frac{\kappa}{2} -i \Delta \Bigl) \langle a \rangle - K \cdot ig \sum_{{m}} f(x_{{m}}) \langle \sigma^\mathrm{ge}_{{m}} \rangle \\
\dot{\langle a^{\dagger} a \rangle}=&- \Bigl( \kappa-2i \Delta \Bigl) \langle a^{\dagger} a \rangle +i g K \sum_{{m}} f(x_{{m}})(\langle a \sigma^\mathrm{eg}_{{m}}\rangle - \langle a^{\dagger}\sigma^\mathrm{ge}_{{m}}) \\
\dot{\langle a a \rangle}=&- \Bigl( \kappa -2i \Delta \Bigl) \langle a a \rangle -2 i g K \sum_{{m}} f(x_{{m}})\langle a \sigma^\mathrm{ge}_{{m}}\rangle  \\
\dot{\langle a \sigma^\mathrm{eg}_{{m}} \rangle} =&-\Bigl( \frac{\gamma+\kappa + \nu}{2} -i\Delta \Bigl) \langle a \sigma^\mathrm{eg}_{{m}} \rangle +ig f(x_{{m}}) \langle a^{\dagger}a \rangle -ig \sum_{{j=1}}^{N_\mathrm{Cl}} \sum_{i=1}^K f(x_{{j}}) \langle \sigma^\mathrm{eg}_{{m}} \sigma^\mathrm{ge}_{{j, i}} \rangle \\[-10 pt]
&-2igf(x_{{m}}) \langle a^{\dagger}a \sigma^{\mathrm{ee}}_{{m}}\rangle \nonumber \\[10 pt]
\dot{\langle a \sigma^\mathrm{ge}_{{m}} \rangle} =& -\Bigl( \frac{\gamma+\kappa + \nu}{2} -i\Delta \Bigl) \langle a \sigma^\mathrm{ge}_{{m}} \rangle -ig f(x_{{m}}) \langle a a \rangle -ig \sum_{{j=1}}^{N_\mathrm{Cl}} \sum_{i=1}^K f(x_{{j}}) \langle \sigma^\mathrm{ge}_{{m}} \sigma^\mathrm{ge}_{{j, i}} \rangle \\[-10 pt]
&+2igf(x_{{m}}) \langle a a \sigma^{\mathrm{ee}}_{{m}}\rangle \nonumber \\[10 pt]
\dot{\langle \sigma^{\mathrm{ee}}_{{m}} \rangle} =&igf(x_{{m}})(\langle a^{\dagger}\sigma^\mathrm{ge}_{{m}} \rangle-\langle a \sigma^\mathrm{eg}_{{m}}\rangle) -(\gamma +\nu)\langle \sigma^{\mathrm{ee}}_{{m}} \rangle + \nu  \\[13 pt]
\dot{\langle \sigma^\mathrm{ge}_{{m}} \rangle}=& -\frac{\gamma +\nu}{2} \langle \sigma^\mathrm{ge}_{{m}} \rangle -igf(x_{{m}})\langle a \rangle +2igf(x_{{m}}) \langle a \sigma^{\mathrm{ee}}_{{m}} \rangle \\[13 pt]
\dot{\langle a \sigma^{\mathrm{ee}}_{{m}} \rangle}=&-(\gamma + \nu) \langle a \sigma^{\mathrm{ee}}_{{m}} \rangle -\frac{\kappa}{2} \langle a \sigma^{\mathrm{ee}}_{{m}} \rangle +\nu \langle a \rangle -igf(x_{{m}})\langle a a \sigma^\mathrm{eg}_{{m}} \rangle  \\[4 pt]
&-ig \sum_{{j=1}}^{N_\mathrm{Cl}} \sum_{i=1}^K f(x_{m})\langle \sigma^{\mathrm{ee}}_{{m}}\sigma^\mathrm{ge}_{{j, i}}\rangle+ig f(x_{{j}})\langle a^{\dagger}a\sigma^\mathrm{ge}_{{m}}  \rangle. \nonumber
\end{align}
\end{subequations}
For the sake of better readability we refrain from carrying out all the expansions of three-operator expectation values like $\langle a a \sigma^\mathrm{ee}_\mathrm{m} \rangle$. In order to end up with a closed set of equations, one needs to approximate the three-operator expectation values by two-operator expectation values according to Eq.~\eqref{eq:3opexpansion}. By neglecting the atom-atom correlations we get rid of the quadratic scaling, but we are still able to keep higher order correlations, namely the ones between atom and field. 

Grouping the atoms into clusters is more involved in the second order and mixed order as opposed to first order, one has to carefully take into account for example the correlations within a cluster~\cite{Bychek2021superradiant, fasser2022master}.

\section*{Determining the steady state photon number}
\begin{figure}[ht]
\includegraphics[width=1.0\columnwidth]{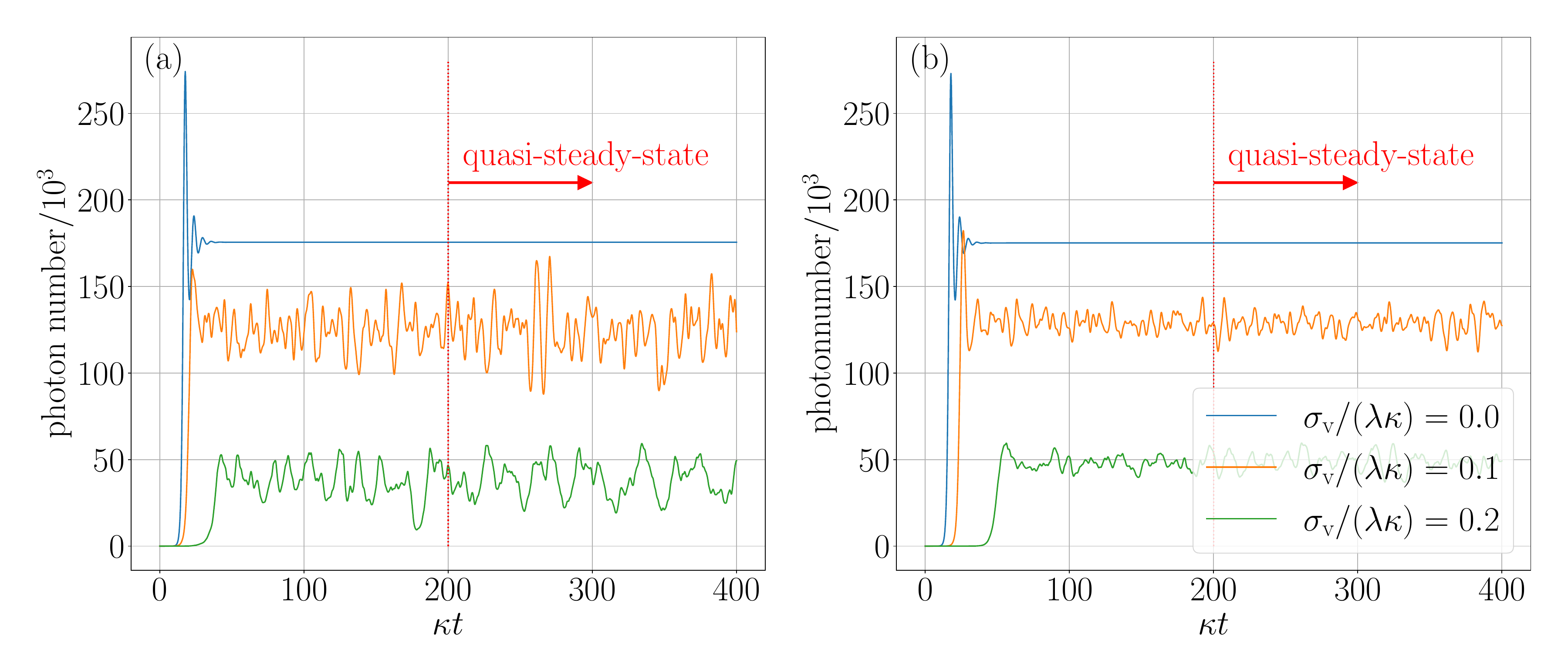}
\caption{\emph{Determining the steady state photon number.} Mean photon number time evolution for different $\sigma_\mathrm{v}$ with (a) 200 atom clusters and (b) 1000 atom clusters. After a certain time, the system reaches its quasi steady state. The fluctuations become smaller if we divide the velocity distribution into a larger number of atom clusters. The other parameters are the same as in~\fref{fig.fixg} for $N = 10^6$.}\label{fig.exampletraj}
\end{figure}
In order to study the photon number values depending on different parameters we initialize the set of differential equations with a certain parameter set and let the system evolve until it reaches its quasi-steady-state. The velocities are drawn from a Gaussian distribution with width $\sigma_\text{v}$ corresponding to a temperature. 
We average over the photon number from the timepoint, at which the quasi-steady-state has been reached, see~\fref{fig.exampletraj}. Using more clusters leads to much smaller fluctuations, however, this also requires more computation time. Finding a compromise between high clusternumbers and reasonable computation time is discussed in the next section. Moreover, one can imagine that certain samples of the distribution exhibit special properties, in order to average them out we use 50 (unless indicated otherwise) different samples and also average over them. 

\section*{Comparison expansion orders and number of clusters}
Limited computation time forces us to make compromises, as both higher expansion orders and higher number of clusters describe the physical reality better, but also require more computation time. 

\begin{figure}[ht]
\includegraphics[width=1.0\columnwidth]{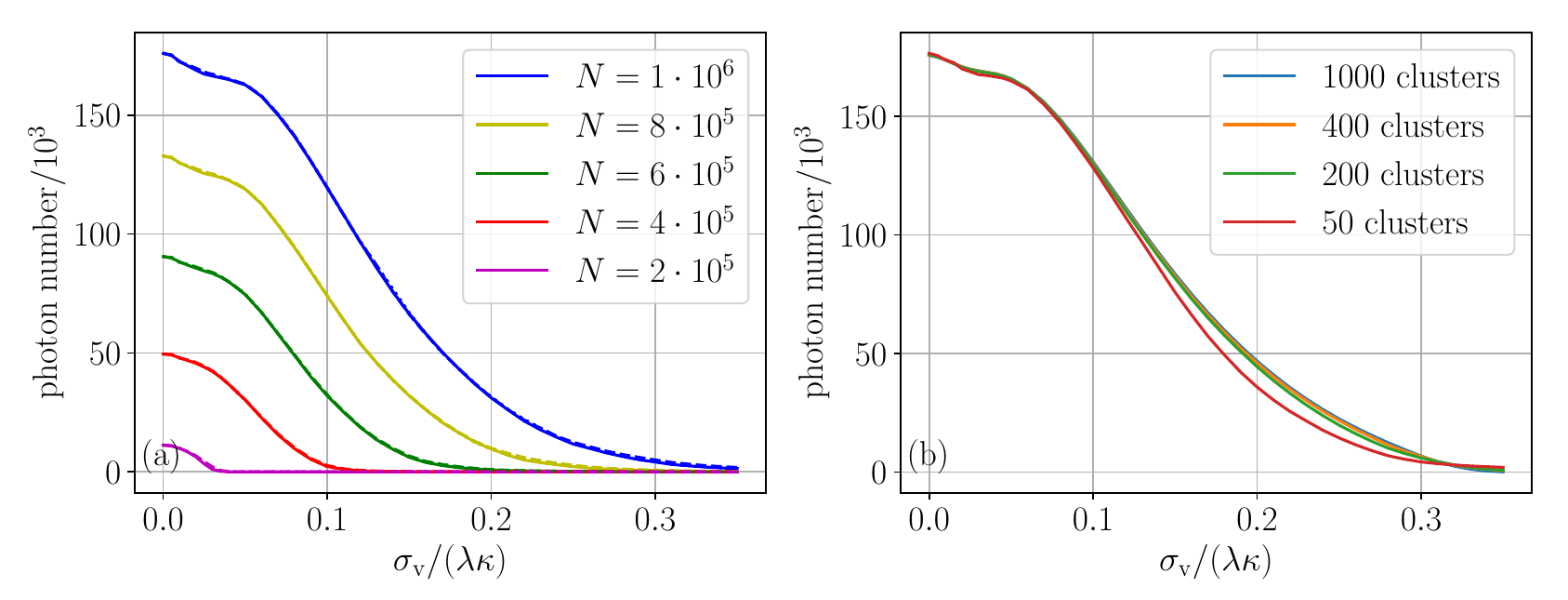}
\caption{\emph{Expansion orders and number of clusters.} (a) The solid line represents mean field, the dotted line is mixed order and the dashed line depicts the result in second order, we use $N_\mathrm{Cl}=25$. The results for all orders agree very well. (b) The chosen number of clusters $N_\mathrm{Cl}$ makes a difference, especially for medium temperatures. However, increasing the number of clusters from $N_\mathrm{Cl} \approx 400$ has only a marginal effect on the results. The other parameters are the same as in~\fref{fig.fixg} for $N = 10^6$.} \label{fig.savcomptime}
\end{figure}

In the following we compare the results for different expansion orders and number of cluster. In~\fref{fig.savcomptime}(a) the solid line depicts the first order, the dotted line mixed order and the dashed line represents the data in second order. We see that all three lines almost perfectly align. Therefore, using higher expansion orders does not lead to significant differences and the first order approximation captures the essential physics in our parameter regime.

In~\fref{fig.savcomptime}(b) we depict the results in first order for different clusternumbers ranging from $N_\mathrm{Cl}=50$ to $N_\mathrm{Cl}=1000$. Here we see, that varying the number of clusters $N_\mathrm{Cl}$ does have an influence on the results. We chose $N_\mathrm{Cl}=400$, as the difference between choosing $N_\mathrm{Cl}=400$ and as high as $N_\mathrm{Cl}=1000$ is only marginal, while the computation time significantly increases.

\section*{Acknowledgements} 

We acknowledge funding from the Austrian Science Fund (FWF) doctoral college No. DK-ALM W1259-N27 (M.F.) and the European Union’s Horizon 2020 research and innovation program under the Grant Agreement No. 820404 iqClock (C. H., D. P., H. R.). 
 
\section*{Data availability} 

\section*{Underlying data}

Figshare: Superradiant$\_$laser$\_$Figures. 
\url{https://doi.org/10.6084/m9.figshare.c.6781920.v1} 
\cite{fasser_2023}.
\\
This project contains the following underlying data:
\begin{itemize}
	\item Data used in Figures 2-6. All data are stored using DelimitedFiles.jl package in Julia.
	\item Data are stored together with the source code (see below), specific files to readout and plot the data are provided.
\end{itemize}

Data are available under the terms of the \href{https://creativecommons.org/publicdomain/zero/1.0/}{Creative Commons Zero "No rights reserved" data waiver} (CC0 1.0 Public domain dedication).

\section*{Software availability}
 
\begin{itemize}
	\item Source code available from: \href{https://github.com/martinf97/SRL\_paper}{https://github.com/martinf97/SRL\_paper}
	\item Archived source code at time of publication:
	\href{https://doi.org/10.5281/zenodo.8232295}{https://doi.org/10.5281/zenodo.8232295} \cite{fasser_2023_git}
	\item License: \href{https://opensource.org/licenses/MIT}{MIT License}
\end{itemize}

{\small\bibliographystyle{ieeetr}}
\bibliography{main}

\end{document}